\documentclass[twoside]{LCWS11}
\usepackage[latin1]{inputenc}
\usepackage{graphicx,epsfig,color}
\usepackage{wrapfig,rotating}
\usepackage{amssymb,amsmath,array}
\usepackage{cite}
\begin{document}
\title{Top Mass Measurement at CLIC at 500 GeV} 

\author{Frank Simon$^{1, 2}$, Katja Seidel$^{1, 2}$, Stephane Poss$^{3}$
\vspace{.3cm}\\
1- Max-Planck-Institut f\"ur Physik,  M\"unchen, Germany
\vspace{.1cm}\\
2- Excellence Cluster `Universe', TU M\"unchen, Garching, Germany
\vspace{.1cm}\\
3- CERN, Geneva - Switzerland
}

\maketitle

\begin{abstract}
We present a study of the capability of a 500\,GeV $e^+e^-$ collider based on CLIC technology for precision measurements of top quark properties. The analysis is based on full detector simulations of the CLIC\_ILD detector concept using Geant4, including realistic background contributions from two photon processes. Event reconstruction is performed using a particle flow algorithm with stringent cuts to control the influence of background. The mass and width of the top quark are studied in fully-hadronic and semi-leptonic decays of $t\bar{t}$ pairs using event samples of signal and standard model background processes corresponding to an integrated luminosity of 100\,fb$^{-1}$. Statistical uncertainties of the top mass given by the invariant mass of its decay products of 0.08\,GeV and 0.09\,GeV are obtained for the fully-hadronic and the semi-leptonic decay channel, respectively,  demonstrating that similar precision to that at ILC can be achieved at CLIC despite less favorable experimental conditions.
\end{abstract}

\section{Introduction}

The top quark plays a unique role in particle physics. Due to its high mass, it is particularly sensitive to new physics and is intimately connected to the mechanism of electroweak symmetry breaking. It has sizable impact on the Higgs boson mass through radiative corrections, and, together with the W boson mass, drives electroweak predictions for the Higgs mass. Due to its short lifetime, the top quark decays before hadronizing, offering the unique opportunity to study a bare quark by accessing its properties directly through its decay products. Top quarks decay electroweakly, into a real $W$ boson and a down-type quark. Due to the large $bt$ CKM matrix element, the decay is almost exclusively into a $W$ boson and a $b$ quark. 

To date, top quarks have been observed at the Tevatron and at the LHC. At present, the best measurement of the mass is provided by the Tevatron, with a statistical error of 0.6\,GeV \cite{Lancaster:2011wr}. The measurement is already systematically limited, with a total systematic error of 0.75\,GeV. Early LHC analyses obtained statistical errors on the order of 1\,GeV to 2\,GeV, with systematic errors close to 3\,GeV \cite{ATLAS-CONF-2011-120, CMS-PAS-TOP-10-009}. With increasing integrated luminosity, significant improvement is anticipated, but the systematics are expected to remain substantial due to the challenging environment of hadron colliders and due to theoretical uncertainties \cite{Aad:2009wy}. 

Significant improvements are expected in $e^+e^-$ collisions, which provide a cleaner experimental environment. The theoretically cleanest way of measuring the top quark mass is by means of a threshold scan, since such measurements can be linked directly to theoretically meaningful mass definitions. Studies suggest  that  combined theoretical and experimental errors of significantly below 100\,MeV can be achieved with this technique \cite{Martinez:2002st}. A drawback of this technique is that a threshold scan requires the collider to be operated for a single measurement over an extended period, conflicting with other studies to be performed at such a machine. It is thus attractive to explore the possibilities for top property measurements in $t\bar{t}$ production well above threshold, by performing the top reconstruction from its decay products, the same technique as used in hadron colliders. Here, the theoretical interpretation of the observations is more challenging, but progress has been made recently in establishing connections between the top mass parameter used in theory and the experimentally observable invariant mass of the decay products \cite{Fleming:2007qr, Fleming:2007xt}. In the following, the invariant mass of the top decay products will be referred to as the top mass.

The relatively clean environment of $e^+e^-$ collisions, combined with the expected jet energy and track momentum resolution of linear collider detectors, makes precision measurements in fully-hadronic ($e^+e^- \rightarrow t\bar{t} \rightarrow q\bar{q}b\,q\bar{q}b$) and semi-leptonic ($e^+e^- \rightarrow t\bar{t} \rightarrow q\bar{q}b\, l\nu b$) decay channels, the channels with the highest branching fraction, possible. For the ILC, studies with full detector simulations have shown that statistical errors on the level of 100\,MeV can be achieved for integrated luminosities of 100\,fb$^{-1}$ at $\sqrt{s}$ = 500\,GeV \cite{:2010zzd, Aihara:2009ad}.

In the framework of the CLIC Conceptual Design Report \cite{CLIC_CDR}, top quark pair production was studied as a benchmark to evaluate the physics performance for processes with multi-fermion final states at 500 GeV, including the performance of flavor tagging in the CLIC environment. This process also provides the possibility for a direct comparison to the detector performance studies performed for the ILC detector letter of intents \cite{:2010zzd, Aihara:2009ad}. 

\section{Experimental Conditions at a 500\,GeV CLIC Collider}

The Compact Linear Collider CLIC \cite{Assmann:2000hg} is a collider concept based on normal conducting accelerating cavities and two-beam acceleration, which is designed to provide up to 3\,TeV collision energy. In a staged approach, a shorter, lower energy version would be operated initially, while construction is under way for the full energy phase.  

Here, we study the case of a 500\,GeV CLIC machine, which is directly comparable to the baseline design of the International Linear Collider. The use of a different acceleration technology leads to differences in the experimental environment which could potentially have a negative impact on the physics performance. The most important difference between ILC and CLIC in that respect is the time between bunch crossings within a bunch train, which is 0.5 ns in the case of CLIC, while it is 356 ns or 670 ns in the case of the ILC, depending on the adopted design. For typical detector integration times of the order of a few to 100 ns, the short bunch crossing time leads to the pile-up of background from many bunch crossings over the 177 ns long bunch trains, which is not present at ILC. In addition, the tighter focusing at CLIC results in increased beamstrahlung and correspondingly larger energy spread, with $\sim$61\% of the total luminosity within 1\% around the peak, compared to $\sim$72\% at the ILC. This translates into larger uncertainties when using energy or momentum constraints along the beam axis.  

The radiated photons lead to background through the creation of coherent and incoherent $e^+e^-$ pairs as well as incoherent quark pair production, which results in hadronic events. While the coherent pairs are emitted at very small angles, defining the crossing angle of the machine, the incoherent pairs have higher transverse momenta and constrain the dimension of the beam pipe in the experiment as well as the radius of the vertex detector. The hadron background affects all aspects of the event reconstruction, in particular jet energy measurements. At a 500 GeV CLIC machine, 0.3 $\gamma\gamma \rightarrow \mathrm{hadrons}$ events per bunch crossing are expected, with an energy of 13.3 GeV. 3.4 GeV of energy are deposited in the calorimeter system, 0.2 GeV out of this in the barrel detectors.

The detector model used in the present study is a variant of CLIC\_ILD \cite{CLIC_ILD},  a detector concept based on Particle Flow event reconstruction. It consists of a low mass, high precision vertex detector and an inner silicon tracker, surrounded by a large-volume time projection chamber, followed by highly granular electromagnetic and hadronic calorimeters contained inside a 4 T solenoidal magnet with instrumented flux return for muon identification. The detector design is based on the ILD detector concept \cite{:2010zzd} for the ILC, adapted to account for the higher energy and more severe background conditions at CLIC \cite{CLIC_CDR}. This leads to an increased radius of the innermost layer of the vertex detector, which sits at 31\,mm compared to 16\,mm in ILD at the ILC. At 500 GeV, the background is significantly reduced compared to the 3 TeV case, permitting modifications of the detector to optimize its performance for the lower collision energy. In particular the innermost vertex detector layer for CLIC\_ILD can move in by 6 mm to a radius of 25\,mm, improving flavor tagging at low momentum.

Minimization of the impact of the hadronic background requires strict timing cuts on the reconstructed particles in the particle flow event reconstruction to limit the influence of out-of-time contributions. Here, timing in the calorimeters is of particular importance.

\section{Event generation, simulation and reconstruction}

The signal process $e^+e^- \rightarrow t\bar{t}$, has a cross section of approximately 530\,fb at a 500 GeV CLIC collider. The top quark decays almost exclusively into a $W$ boson and a $b$ quark. The signal events can thus be grouped into different classes, according to the decay of the $W$ bosons. These are the {\it fully-hadronic} channel, with both $W$s decaying into quark pairs, the {\it semi-leptonic} channel, with one $W$ decaying into quarks, the other into a lepton and the corresponding neutrino, and the {\it fully-leptonic} channel, with both $W$s decaying into lepton and neutrino. In the leptonic channels, the decay into a $\tau$ and a neutrino is a special case, since the $\tau$ decays almost instantly into either a lepton and a neutrino or into one or more hadrons and a neutrino, giving rise to additional missing energy in the final state, and potential confusion with hadronic decay modes. 

In the present analysis, only fully-hadronic and semi-leptonic events, excluding $\tau$ final states, were selected, since those provide the best possible mass measurement. However, to account for imperfect event classification, all possible decay modes of the $t\bar{t}$ pair were generated according to their respective branching fractions. The top mass and width, as defined in the event generator, were fixed for the signal event sample to $m_{t}$ = 174.0\,GeV and $\sigma_{t}$ = 1.37\,GeV.

In addition to the signal, background processes with similar event topologies have to be considered. These are mostly four and six fermion final states, with the high cross-section quark pair production in addition. Beyond this, the processes  $q \bar q e^+ e^-$ and $q \bar q e \nu$, which are dominated by t-channel single boson production, were investigated using samples with reduced statistics. It was shown that the non-di-boson contributions are rejected completely in the analysis. Since the di-boson contributions are accounted for in the $e^+e^-\rightarrow WW$ and  $e^+e^-\rightarrow ZZ$ modes, the processes $q \bar q e^+ e^-$ and $q \bar q e \nu$ were not considered in the final production.

\begin{table}
\centering
\begin{tabular}{c||c|c|c}
\multicolumn{4}{c}{$\sqrt{s} =$ 500\,GeV, CLIC beam energy spectrum}\\
\hline
process type & $e^+ e^- \rightarrow$ &  cross section $\sigma$ &event generator \\
\hline
\hline
Signal ($m_{\textrm{t}}$ = 174\,GeV)& $t \bar t$	& 528\,fb	& PYTHIA	 \\
\hline
Background& $W W$	& 7.1\,pb	& PYTHIA	 \\
Background& $Z Z$	& 410\,fb	& PYTHIA	 \\
Background& $q \bar q$	& 2.6\,pb 	& WHIZARD	 \\
Background& $W W Z$	& 40\,fb		& WHIZARD	\\

\end{tabular} 
\caption{Signal and considered background processes, with cross sections calculated for CLIC500.\label{tab:processes}}
\label{tab:channel-production}
\end{table}

Since WHIZARD 1.95 \cite{Kilian:2007gr}, which was used as the default event generator for the CLIC CDR benchmark studies \cite{CLIC_CDR}, is not accurately reproducing final-states and explicitly defined intermediate states with particles with non-zero width, PYTHIA \cite{Sjostrand:2006za} was used to generate the signal process $e^+e^-\rightarrow t\bar{t}$ as well as the two background processes $e^+e^-\rightarrow WW$ and  $e^+e^-\rightarrow ZZ$. The processes with explicitly given final states, without specifying intermediate particles, $e^+e^-\rightarrow q \bar q$, $e^+e^-\rightarrow q\bar{q} e^+e^-$ and $e^+e^-\rightarrow q\bar{q} e \nu$ were generated with WHIZARD 1.95. Since the process $e^+e^-\rightarrow WWZ$ is not implemented in PYTHIA, WHIZARD was used for its generation. For simplicity, these events were generated with zero width for the intermediate bosons, allowing to specify defined intermediate states in WHIZARD, keeping integration times short. Table \ref{tab:processes} summarizes the studied processes and the corresponding event generators, with approximate cross sections at a 500 GeV CLIC machine.

All generated events are fully simulated in a detailed detailed GEANT4 \cite{Agostinelli:2002hh} model of the CLIC\_ILD detector introduced above. Before reconstruction, all events are overlayed with $\gamma\gamma \rightarrow$ hadrons background events corresponding to 300 bunch crossings, the equivalent of a full bunch train. In the particle flow event reconstruction with the PandoraPFA algorithm \cite{Thomson:2009rp}, $p_t$ dependent timing cuts are applied to reduce the impact of this background on the signal. To keep the loss of signal particles in the 500 GeV events, which are characterized be relatively low energy particles, at a minimum, the timing cuts are considerably relaxed compared to the standard CLIC 3 TeV event reconstruction. Due to the significantly lower background multiplicity, a clean reconstruction of the signal processes is still possible.

\section{Analysis and Results}

The goal of the analysis is to provide a high-purity $t \bar t$ event sample with well reconstructed events to achieve the best possible measurement of the invariant mass of the top quark decay products.  It favors strict rejection of imperfectly reconstructed events over the maximization of reconstructed top quark candidates. In general, the present analysis scheme is similar to the $t \bar t$ analysis performed for the ILD Letter of Intent. Due to a more general input sample in the present study, which includes semi-leptonic $\tau$ events as well as fully-leptonic events, and due to the different bunch and beam structure of the CLIC machine, some additional analysis steps had to be introduced, while major strategy changes had to be adopted for other parts of the analysis. 

The first step of the analysis is the classification of the events into three branches according to the number of isolated leptons (electron or muon) identified in the event. Events without isolated leptons are classified as fully-hadronic, events with one or more leptons are classified as semi-leptonic and fully-leptonic, respectively. The fully-leptonic events are rejected, while events in the other two classes are clustered into jets for the subsequent analysis, using an exclusive $k_t$ algorithm with $\Delta \eta$, $\Delta \phi$ metric as implemented in the FastJet package \cite{Cacciari:2005hq} and a jet size parameter of $R$ = 1.3, selected to account for the rather large jets at relatively low energy. Fully-hadronic event candidates are clustered into six jets, while semi-leptonic events are clustered into four jets, with the identified isolated lepton excluded from jet-finding. 

Following this, flavor tagging of the found jets is performed using the LCFI Flavour Tagging  package. It is based on a neural network, which provides $b$ and $c$ jet probabilities (``b-tag'') for each jet in the event. The tagging is crucial for the correct assignment of jets to top candidates in $t \bar t \rightarrow (bq\bar q) (\bar b q\bar q)$ and $t \bar t \rightarrow (bq\bar q)(\bar b l \nu_l)$ events and provides discrimination of signal events and multi-fermion background. To construct the final state, the two jets with the highest b-tag are identified as the two b-quarks from the decay of the two top quarks. In the semi-leptonic case, one of the $W$ bosons is constructed from the two remaining jets, the other from the isolated lepton and the neutrino, measured via missing momentum. In the all-hadronic case, a three-fold ambiguity exists in the assignment of light-flavor jets to $W$ bosons. Here, the combination with the smallest overall difference of the invariant mass of the two $W$ candidates from the true $W$ mass is selected.

After the identification of $b$ jets and the pairing of light jets and leptons into $W$ bosons, the next step is the grouping of $W$ candidates and $b$ jets into top quarks. This assignment is  performed with a kinematic fit. Out of the two possible combinations, the one with the higher probability of the kinematic fit result is chosen as the correct combination of $W$ bosons and $b$ jets into top candidates. The kinematic fit, here taken from the MarlinKinFit package \cite{MarlinKinFit}, uses kinematic constraints assuming a $t \bar t$ event to improve the precision of the event parameters of interest, in the present case the invariant mass of the top quark candidate.

In this analysis, the input parameters to the kinematic fit are the four-momenta of the light jets, already paired into $W$ bosons in the case of the 6-jet sample, the momenta of the two $b$ jets, and that of the isolated lepton for semi-leptonic events. In the latter case, the unmeasured neutrino represented an invisible particle in the fit, with starting values set to the measured missing energy and momentum in the event.  In the fit, the energy of the system is constrained to the nominal center of mass energy, not accounting for the beam energy spectrum. This leads to the rejection of events with particularly low center of mass energy, but overall significantly improves the reconstruction of the invariant mass. Additional constraints are momentum conservation, accounting for the beam crossing angle, the mass of the intermediate $W$ bosons, and  the requirement for an equal mass of the $t$ and the $\bar t$ candidate.

During the fit procedure, the fitter varies the particle momenta and energies to fulfill the constraints. This is done within the specified detector resolution for the various input particles, both in energy and azimuthal and polar angle. The angular resolutions for jets and angular and energy resolutions for leptons is derived from Monte Carlo studies of a $t \bar t$ sample, while the energy resolution for jets is taken from the single jet performance of PandoraPFA, in order to avoid an overestimation of the error due to correlated effects between several jets originating from the same object.

\begin{figure}[!ht]
\centering
  \includegraphics[width=0.8\textwidth]{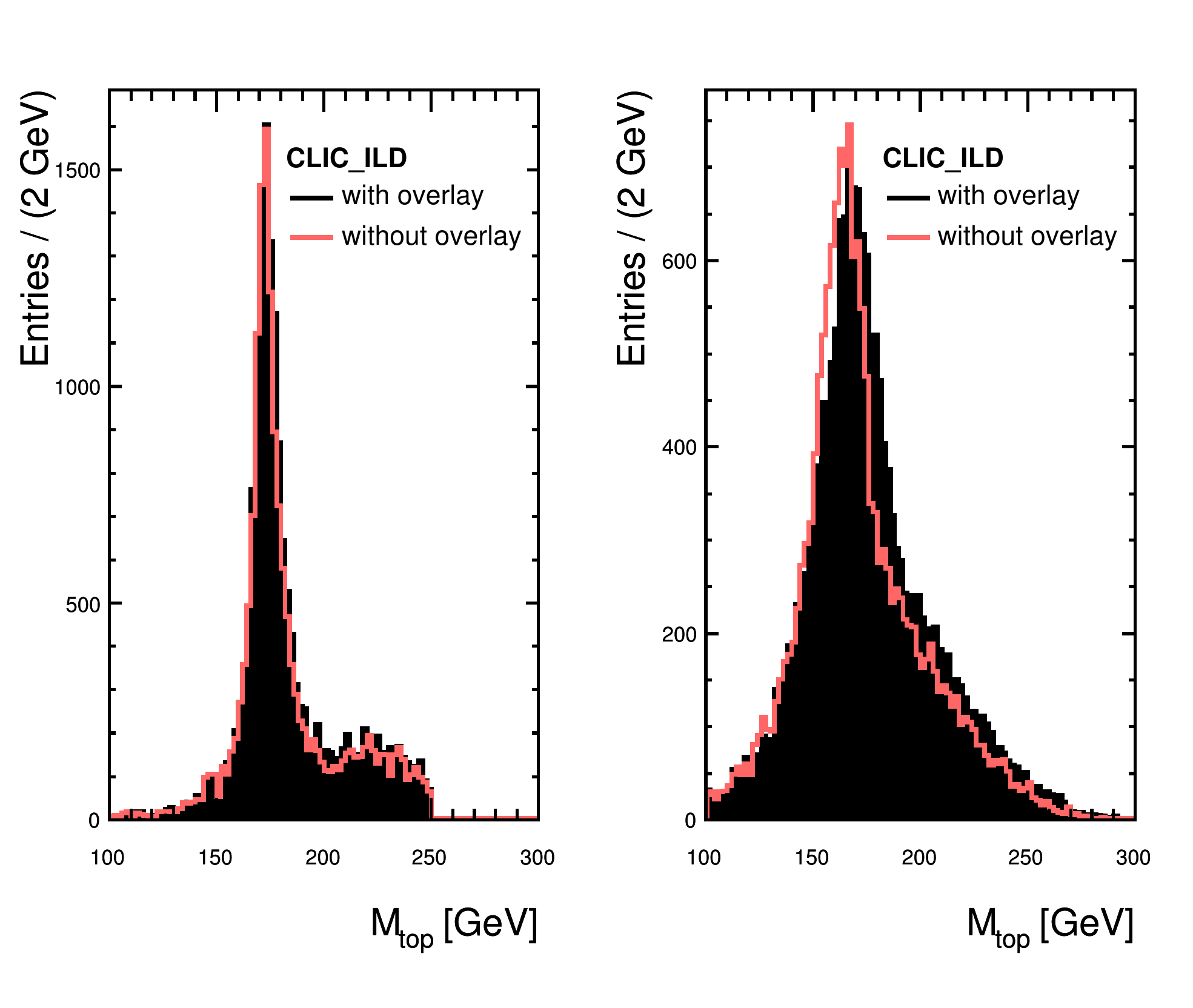}
	\caption{Reconstructed invariant mass distribution for the 6-jet channel without (right) and with (left) kinematic fit for signal-only events. The black lines shown the top mass distribution for events with overlayed $\gamma \gamma \rightarrow$ hadrons events, the red line shows events without. The high-mass tail in the mass distribution obtained with kinematic fit is due to kinematic reflections in events with incorrect assignment of jets to top candidates.}
   \label{fig:KinFit-FullHad}
 \end{figure}

The fit fails if it is unable to satisfy the constraints outlined above within the allowed modifications of the input parameters. It is observed that some of the fit failures are due to the wrong identification of one of the $b$ jets. This is particularly likely in the case of a $W$ decaying into a charm quark and another light quark. Thus, to improve the number of successful fits and to account for possible wrong flavor tagging, the kinematic fit is repeated for unsuccessful kinematic fits after exchanging the b-jet with the lower b-tag value with the light-jet with the highest b-tag value. This procedure increases the number of successful kinematic fits by $\sim 20\%$.

The result of the kinematic fit, compared to the top mass reconstruction without kinematic fit, is shown in Figure \ref{fig:KinFit-FullHad} for the fully-hadronic signal-only event sample. Also shown is the comparison of the performance with and without the inclusion of $\gamma \gamma \rightarrow$ hadrons background. It is apparent that the kinematic fit significantly improves the reconstructed invariant mass. In addition, it also reduces the impact of the of $\gamma \gamma \rightarrow$ hadrons background.

Beyond the improvement of the mass measurement, the kinematic fit also serves as an excellent tool for the rejection of non-$t\bar t$ background, as well as for the suppression of remaining contributions from wrongly classified all-leptonic decays and semi-leptonic events with $\tau$ final states. Additional background rejection is provided by a selection algorithm based on a binned likelihood technique, using event shape, flavor tagging and reconstructed mass parameters. Together with the kinematic fit, the background contaminations are reduced to a level of 11\% for the fully-hadronic and 5\% for the semi-leptonic branch. The overall signal selection efficiencies for the complete analysis are 35\% for fully-hadronic and 56\% for semi-leptonic events. The higher selection efficiency for semi-leptonic top pair decays is mainly due to the weaker constraints in the kinematic fit due to the missing energy and momentum of the neutrino. This results in a slightly broader invariant mass distribution.

 \begin{figure}
	\centering
		\includegraphics[width=0.5\textwidth]{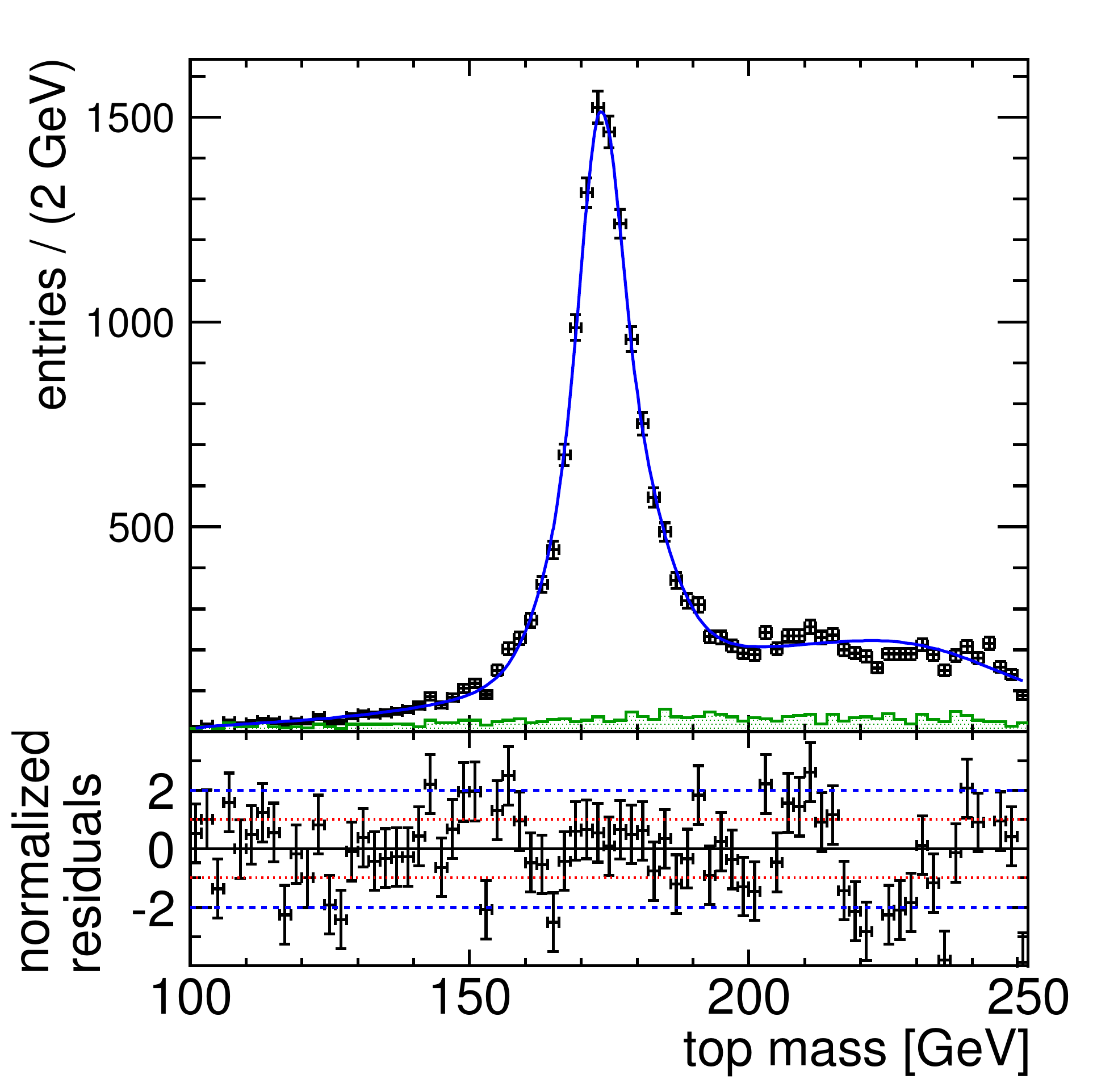}\hfill	
		\includegraphics[width=0.5\textwidth]{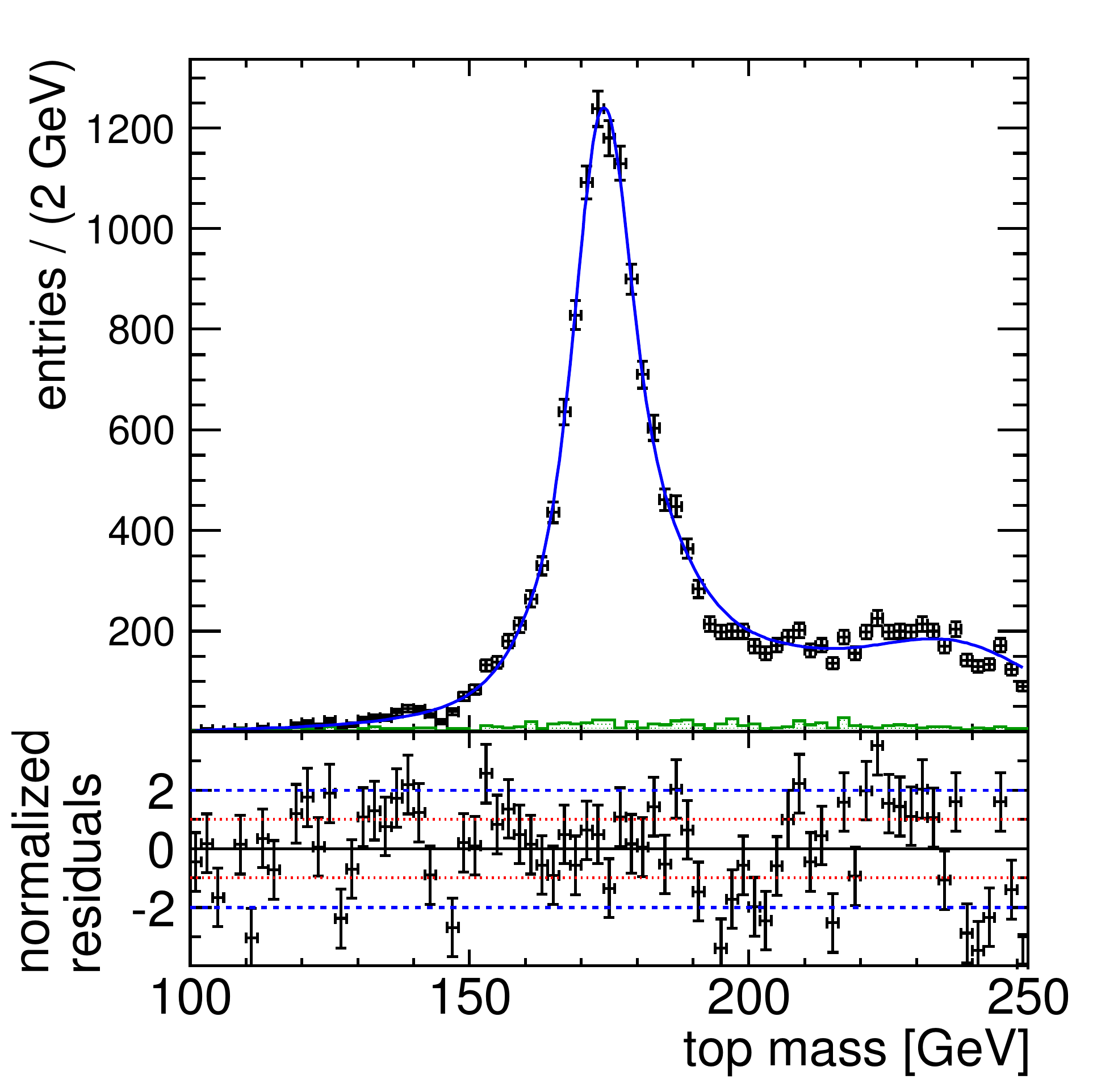}
	\caption{Final invariant top mass distribution for 6-jet events ({\it left}) and 4-jet events ({\it right}). Black points with error bars indicated simulated data classified at signal events. The green hatched histogram shows the contribution of non $t\bar t$ background. The blue line indicates the fit of the top mass.}
  \label{fig:fit-final}
\end{figure}

Figure \ref{fig:fit-final} shows the final reconstructed distribution of the invariant mass of the top decay products, together with contributions from non $t \bar t$ background, for the fully-hadronic and the semi-leptonic event sample. From these distributions, the top mass and width is determined with an un-binned maximum likelihood fit, which includes a Breit-Wigner distribution for the mass and width, convoluted with a detector resolution function, as well as a background contribution and a component accounting for incorrectly reconstructed events ending up in the high-mass region of the distribution. The detector resolution function is determined from a higher statistics fully simulated signal-only sample, and was cross-checked for different values of the top mass and width. 

The resulting top mass is
\begin{equation*}
m_{top} = 174.07\,\mathrm{GeV} \pm 0.08\,\mathrm{GeV (stat)}
\end{equation*}
for the all-hadronic sample, and
\begin{equation*}
m_{top} = 174.28\, \mathrm{GeV} \pm 0.09\,\mathrm{GeV (stat)}    
\end{equation*}
for the semi-leptonic sample. The generated top mass is 174,GeV, thus the all-hadronic mass is in excellent agreement with the input value, while the semi-leptonic measurement differs by three standard deviations. This deviation is likely mostly due to uncertainties of the detector resolution function, which was determined from a statistically independent sample of approximately 2.5 times the integrated luminosity of the signal sample. It is expected that this can be improved with a more thorough study. 

Also the width of the top quark was extracted from the signal fit. For the all-hadronic sample, a width of 
\begin{equation*}
 \sigma_{top} = 1.33\,\mathrm{GeV} \pm 0.21\,\mathrm{GeV (stat)} 
 \end{equation*}
 was obtained, to be compared with a generator level value of 1.37,GeV. For the semi-leptonic sample, a width of
 \begin{equation*}
 \sigma_{top} = 1.55\,\mathrm{GeV} \pm 0.26\,\mathrm{GeV (stat)} 
 \end{equation*}
 is found, also in good agreement with the generator value.

\section{Conclusions}

The mass of the top quark is one of the key parameters of the standard model, and provides sensitivity to new physics. The present study, using full simulations including machine and physics backgrounds, carried out in the framework of the CLIC CDR, shows that a 500 GeV linear $e^+e^-$ collider based on CLIC technology is an excellent tool for precision top measurements. The $\gamma\gamma \rightarrow {\rm hadrons}$ background can be controlled by timing cuts and by a suitable choice of the jet finder, and does not significantly affect the flavor tagging for top events. Precise reconstruction of the event kinematics is achieved by means of a kinematic fit, which also serves to control the energy uncertainty due to the beam energy spectrum and contributed to the rejection of non-$t\bar{t}$ background. With an integrated luminosity of 100\,fb$^{-1}$, a statistical of precision of 80 MeV is achieved in the fully-hadronic decay channel, and a precision of 90 MeV was achieved in the semi-leptonic channel. This precision is comparable to that expected for the ILC, despite the more challenging experimental environment at CLIC, demonstrating the capabilities for precision measurements at a sub-TeV $e^+e^-$ collider based on CLIC technology.

\begin{footnotesize}

% ----------------------------------------------------------------------------

\end{footnotesize}

\end{document}